\documentstyle[twoside,11pt]{article}

\catcode`\@=11
\long\def\@makefntext#1{
\protect\noindent \hbox to 3.2pt {\hskip-.9pt
$^{{\tenrm\@thefnmark}}$\hfil}#1\hfill}         

\def\thefootnote{\fnsymbol{footnote}}
\def\@makefnmark{\hbox to 0pt{$^{\@thefnmark}$\hss}}    

\def\ps@myheadings{\let\@mkboth\@gobbletwo
\def\@oddhead{\hbox{}
\rightmark\hfil\tenrm\thepage}
\def\@oddfoot{}\def\@evenhead{\tenrm\thepage\hfil
\leftmark\hbox{}}\def\@evenfoot{}
\def\sectionmark##1{}\def\subsectionmark##1{}}

\renewcommand{\thefootnote}{\fnsymbol{footnote}}

\newcounter{sectionc}\newcounter{subsectionc}\newcounter{subsubsectionc}
\renewcommand{\section}[1] {\vspace{25pt}\addtocounter{sectionc}{1}
\setcounter{subsectionc}{0}\setcounter{subsubsectionc}{0}\noindent
    {\twelvebf\thesectionc.\kern0.35cm #1}\par\vspace{8pt}}
\renewcommand{\subsection}[1] {\vspace{25pt}\addtocounter{subsectionc}{1}
    \setcounter{subsubsectionc}{0}\noindent
    {\twelvebf\thesectionc.\thesubsectionc\kern0.35cm #1}\par
    \vspace{8pt}}
\renewcommand{\subsubsection}[1] {\vspace{25pt}\addtocounter{subsubsectionc}{1}
    \noindent
    {\twelverm\thesectionc.\thesubsectionc.\thesubsubsectionc\kern0.35cm
    {\kern1pt\twelveit #1}}\par\vspace{8pt}}

\newcommand{\smalllineskip}{\baselineskip=11pt}

\def\ninecirc{
\begin{picture}(0,0)
\put(4.4,1.8){\circle{7.45}}
\end{picture}}
\def\ninecopyright{\ninecirc\kern2.75pt\hbox{\eightrm c}}

\newcommand{\copyrightheading}[1]
    {\vspace*{-1cm}\baselineskip=11pt{\flushleft
    {\ninerm Fractals, #1}\\
    {\ninerm $\ninecopyright$\,\,\, World Scientific
     Publishing Company}\\
     }}

\def\abstracts#1#2#3{{
    \centering{\begin{minipage}{5.0in}\tenrm\baselineskip=12pt
        \centerline{\twelvebf Abstract}
    \vspace{5pt}
    \parindent=0pc #1\par
    \parindent=1pc #2\par
    \parindent=1pc #3
    \end{minipage}}\par}}

\def\ARTICLES{\kern6.15cm\hbox{${\vcenter{\vbox{
    \hrule height 0.4pt width 4.562truein       
    \hbox{\vrule width 0.4pt            
    height 0.6truein                
    \raise0.565cm\hbox{\kern1pc
    \seventeenbf Articles}}
    \hrule height 0.4pt width 4.562truein}}}$}}  

\renewenvironment{thebibliography}[1]
    {\frenchspacing
     \tenrm\baselineskip=12pt
     \begin{list}{\arabic{enumi}.}
    {\usecounter{enumi}\setlength{\parsep}{0pt}
     \setlength{\leftmargin 12.7pt}{\rightmargin 0pt} 
     \setlength{\itemsep}{0pt} \settowidth
    {\labelwidth}{#1.}\sloppy}}{\end{list}}

\newcounter{itemlistc}
\newcounter{romanlistc}
\newcounter{alphlistc}
\newcounter{arabiclistc}

\newcommand{\fcaption}[1]{
        \refstepcounter{figure}
        \setbox\@tempboxa = \hbox{\footnotesize{\bf Fig.~\thefigure\phantom{00}}#1}
        \ifdim \wd\@tempboxa > 6in
           {\begin{center}
        \parbox{6in}{\footnotesize\smalllineskip{\bf Fig.~\thefigure\phantom{00}}#1}
            \end{center}}
        \else
             {\begin{center}
             {\footnotesize{\bf Fig.~\thefigure\phantom{00}}#1}
              \end{center}}
        \fi}

\newcommand{\tcaption}[1]{
        \refstepcounter{table}
        \setbox\@tempboxa = \hbox{\footnotesize\bf Table~\thetable\phantom{00}#1}
        \ifdim \wd\@tempboxa > 6in
           {\begin{center}
        \parbox{6in}{\footnotesize\smalllineskip\bf Table~\thetable\phantom{00}#1}
            \end{center}}
        \else
             {\begin{center}
             {\footnotesize\bf Table~\thetable\phantom{00}#1}
              \end{center}}
        \fi}

\def\@citex[#1]#2{\if@filesw\immediate\write\@auxout
    {\string\citation{#2}}\fi
\def\@citea{}\@cite{\@for\@citeb:=#2\do
    {\@citea\def\@citea{,}\@ifundefined
    {b@\@citeb}{{\bf ?}\@warning
    {Citation `\@citeb' on page \thepage \space undefined}}
    {\csname b@\@citeb\endcsname}}}{#1}}

\newif\if@cghi
\def\cite{\@cghitrue\@ifnextchar [{\@tempswatrue
    \@citex}{\@tempswafalse\@citex[]}}
\def\citelow{\@cghifalse\@ifnextchar [{\@tempswatrue
    \@citex}{\@tempswafalse\@citex[]}}
\def\@cite#1#2{{$\null^{#1}$\if@tempswa\typeout
    {IJCGA warning: optional citation argument
    ignored: `#2'} \fi}}

\def\fnt#1#2{\footnotetext{\kern-.3em
    {$^{\mbox{\sevenrm #1}}$}{#2}}}

\def\runninghead#1#2{\protect\pagestyle{myheadings}
\markboth{\protect\nineit\,\,\,\,\,#1\hfill}
{\hfill\protect\nineit #2\,\,\,\,\,}}
\font\seventeenbf=cmbx10      scaled\magstep3

\font\twelverm=cmr10      scaled\magstep1
\font\twelveit=cmti10     scaled\magstep1
\font\twelvebf=cmbx10     scaled\magstep1

\font\tenrm=cmr10
\font\tenit=cmti10
\font\tenbf=cmbx10

\font\ninerm=cmr9
\font\nineit=cmti9

\font\eightrm=cmr8

\font\sevenrm=cmr7

\catcode`@=11                   
\def\ps@plain{\let\@mkboth\@gobbletwo
     \def\@oddhead{}\def\@oddfoot{\ninerm\hfil\thepage
     \hfil}\def\@evenhead{}\let\@evenfoot\@oddfoot}

\def\ps@myheadings{\let\@mkboth\@gobbletwo  
\def\@oddhead{\hbox{}
\rightmark\hfil\ninerm\thepage}
\def\@oddfoot{}\def\@evenhead{\ninerm\thepage\hfil
\leftmark\hbox{}}\def\@evenfoot{}
\def\sectionmark##1{}\def\subsectionmark##1{}}
\catcode`@=12

\begin{document}

\runninghead{Fractal Dimensions for Critical Potts Clusters}
{Fractal Dimensions for Critical Potts Clusters}
\renewcommand{\thefootnote}{\fnsymbol{footnote}}      

\thispagestyle{plain}
\setcounter{page}{1}

\copyrightheading{Vol.~0, No.~0 (0000)}

\vspace{6pc}
\leftline{\phantom{\ARTICLES}\hfill}

\vspace{3pc}
\leftline{\hskip-0.1cm\vbox{\hrule width6.99truein height0.15cm}\hfill}

\vspace{2pc} \centerline{\seventeenbf FRACTAL DIMENSIONS}
\baselineskip=20pt \centerline{\seventeenbf  AND CORRECTIONS TO
SCALING} \baselineskip=20pt \centerline{\seventeenbf FOR CRITICAL
POTTS CLUSTERS} \vspace{0.27truein} \centerline{AMNON AHARONY}
\baselineskip=12.5pt \centerline{\it School of Physics and
Astronomy} \centerline{\it Raymond and Beverly Sackler Faculty of
exact Sciences} \centerline{\it Tel Aviv University, Tel Aviv
69978, Israel} \vspace{0.08truein} \centerline{JOONAS ASIKAINEN}
\baselineskip=12.5pt \centerline{\it Laboratory of Physics,
Helsinki University of Technology} \centerline{P. O. Box 1100,
FIN-02150 HUT, Espoo, Finland} \vspace{0.36truein}
\abstracts{Renormalization group and Coulomb gas mappings are used
to derive theoretical predictions for the corrections to the
exactly known asymptotic fractal masses of the hull, external
perimeter, singly connected bonds and total mass of the
Fortuin-Kasteleyn clusters for two-dimensional $q$-state Potts
models at criticality. For $q=4$ these include exact logarithmic
(as well as $\log\log$) corrections. }{}{}

\vspace{0.78truein}
\baselineskip=14.5pt
\section{INTRODUCTION}

\noindent
$q$-state Potts models, with
interaction $-J \delta_{\sigma_i \sigma_j}$ ($\sigma_{i,j}=1,2,...,q$)
for the nearest neighbor (nn) sites $i,j$,
 have played an important role in
condensed matter physics\cite{Wu82}. Here we study geometrical
aspects of the critical Potts clusters, in two dimensions. For an
arbitrary configuration of Potts states, one creates bonds between
neighboring sites which have the same state, $\sigma_i=\sigma_j$,
with a probability $p=1-\exp{(-J/kT)}$. No bonds are created
between sites with $\sigma_i \ne \sigma_j$. Here we study the
fractal geometry at $T_c$ of the clusters, made of sites connected
by bonds\cite{For72}. Specifically, we measure the fractal
dimensions $D_M,~D_H,~D_{EP},~{\rm{and}}~D_{SC}$ describing the
scaling of the cluster's mass, hull, external accessible
perimeter\cite{GA} and singly connected bonds\cite{conig82},
respectively, with its radius of gyration $R$. As emphasized by
Coniglio\cite{coniglio}, many of these fractal dimensions have
been derived analytically\cite{Sal87}. Some others have been found
more recently\cite{Aiz99,Dup00}. These theoretical values are
summarized in Table 1, in terms of the Coulomb gas coupling
constant $g={4 \over \pi} \arccos(-{\sqrt{q} \over 2})$.

\begin{table}[h]
\centering
\begin{minipage}{8.5 cm}
\renewcommand{\arraystretch}{1.2}
\caption{Exact theoretical predictions.}
\vspace{.4cm}
{\begin{tabular}{c c c c c c c}
 &$D_S$ & $q=1$ & $q=2$ & $q=3$ & $q=4$ & $c_S/a$ \\
\hline $g$ & & ${8 \over 3}$&3&${10 \over 3
}$& 4 & \\
\hline $M$ & $(g+2)(g+6)/(8g)$ & ${91 \over 48}$&${15 \over
8}$&${28 \over 15}$&${15
\over 8}$&${1 \over 16}$ \\
$H$ & $1+2/g$& ${7 \over 4}$&${5 \over 3}$&${8 \over 5}$&${3 \over 2}$&
$-{1 \over 4}$\\
${EP}$& $1+g/8$ & ${4 \over 3}$&${11 \over 8}$&${17 \over 12}$&${3 \over 2}$& ${
1 \over 4}$\\
${SC}$ & $(3g+4)(4-g)/(8g)$& ${3 \over 4}$&${13 \over 24}$&${7 \over 20}$&$0$& $
-1$\\
\hline
$\theta$ & $4(4-g)/g$ & 2 & ${4 \over 3}$ & ${4 \over 5}$ & 0 (log) & \\
$\theta'$ & $4/g$ & ${3 \over 2}$ & $ {4 \over 3}$&${6 \over 5}$&1& \\
$\theta''$ & $2/g$ & ${3 \over 4}$ & $ {2 \over 3}$&${3 \over
5}$&${1 \over 2}$ &
 \\
\hline
\end{tabular}}
 \label{TableI}
\end{minipage}
\end{table}

We are currently studying the geometry of such Potts critical
clusters, using numerical Monte Carlo simulations\cite{bbm}. These
simulations show that the asymptotic power law dependence of the
various masses on $R$ is approached relatively slowly, and
therefore the analysis of the data must include {\it correction
terms}, particularly as $q$ approaches 4. The present paper
contains a brief summary of analytic results for several of these
corrections. Our predictions for the correction exponents (apart
from the analytic ones) are also listed in Table 1.

\section{RENORMALIZATION GROUP}
\noindent The first correction relates to the {\it dilution field}
$\psi$, which is generated under renormalization even when one
starts with the non-dilute case\cite{cardy}.
For our non-dilute case, one expects $\psi(\ell)$ to increase
under the renormalization group recursion relations (RGRR's) from
$-\infty$ towards its critical value $\psi^*$ ($\le 0$ for $q \le
4$), while the cluster linear size, like all other lengths,
rescales as $R(\ell)=Re^{-\ell}$. Following Cardy {\it et
al.}\cite{cardy}, we assume that $\ell_0$ prefacing iterations
bring $\psi$ from $-\infty$ up to $\psi(\ell_0)=\psi_0$, with
$|\psi_0| \ll 1$. We then expand the RGRR for $\psi$ in powers of
$\psi$ and $\epsilon=q-4$,
\begin{equation}
\frac{d\psi}{d\ell}=a(\epsilon+\psi^2+b\psi^3+r\epsilon \psi+...).
\label{psi}
\end{equation}

At $q=4$, this yields
\begin{equation}
a(\ell-\ell_0) \approx
\psi_0^{-1}-\psi(\ell)^{-1}+b\log\frac{b+\psi(\ell)^{-1}}
{b+\psi_0^{-1}}.
\label{psil}
\end{equation}
Iterating up to $R(\ell^\times)=1$, we can express
$\psi(\ell^\times)$ in terms of $\log R= \ell^\times$. For large
$R$, $\psi_0/\psi(\ell^\times) \approx A (\log R + B \log(\log
R)+E) +O(\log \log R/\log R)$, where $A=-a \psi_0,~B=-b/a$ and $E$
also depends on $a,~b,~\ell_0$ and $\psi_0$.

For $q<4$, an expansion to second order in
$\epsilon'=\sqrt{-\epsilon}$ yields $\psi(\ell^\times) \approx
\psi^*+ {\tilde B}R^{-\theta}$, with $\psi^*=-\epsilon'(1-(r-b)
\epsilon'/2)+\ldots$, $\theta=2a\epsilon'(1-b \epsilon')+\ldots$
and ${\tilde B} \propto (\psi_0-\psi^*)$.  To order-$\epsilon'$,
one obtains a full solution,
\begin{equation}
\psi(\ell)=-\epsilon' \frac{1+\hat B e^{-\theta \ell}}{1-\hat B
e^{-\theta \ell}}, \label{psilq}
\end{equation}
where $\hat B=(\psi_0+\epsilon')/(\psi_0-\epsilon')$. Indeed,
$\psi$ approaches $\psi^*$ for large $\ell$.

To obtain the scaling of $M_S(R)$, we write the RGRR for the field
$h_S$ conjugate to the density $\rho_S \equiv M_S/R^d$ as
\begin{equation}
\frac{dh_S}{d\ell}=(y_S+c_S\psi(1+e_S\psi+f_S\psi^2+\ldots))h_S,
\end{equation}
where the coefficients may depend on $\epsilon$. $\rho_S$ is then
found as a derivative of the free energy with respect to $h_S$.
For $q=4$, its singular part becomes
\begin{eqnarray}
\rho_S(\ell) & \propto & e^{-d\ell} h_S(\ell)/h_0 \\ \nonumber & =
& \exp[(y_S-d) \ell+\int_{\ell_0}^{\ell}(c_S\psi(1+e_S\psi+f_S
\psi^2+\ldots) )d\ell] \\ \nonumber & \propto & e^{(y_S-d)
\ell}[\psi(\ell)/\psi_0]^{c_S/a} (1+O(\psi(\ell))).
\end{eqnarray}
For large $\log R=\ell^\times$, this becomes
\begin{equation}
M_S \propto R^{D_S}(\log R + B \log(\log R)+ E)^{-c_S/a}
(1+O(\log\log R/\log R)), \label{log}
\end{equation}
with $D_S=y_S(q=4)$, and $c_S/a$ is to be taken from Table 1 (see
below) . Note that $B=-b/a$ is universal (i. e. independent of
$\psi_0$), and the non-universal constant $E$ is the same for all
$S$. Equation (\ref{log}) generalizes the logarithmic corrections
of Cardy {\it et al.}\cite{cardy}.

In practice, the numerical results are always analyzed by looking
at the local logarithmic slope,
\begin{eqnarray}
D_S^{\rm eff}&=&d\log M_S/d\log R=d\log
h_S/d\ell|_{\ell=\ell^\times}
\nonumber\\
&=&y_S+c_S\psi(\ell^\times)(1+e_S \psi(\ell^\times)+f_S
\psi(\ell^\times)^2+\ldots).
\end{eqnarray}
In some cases, this expression (in which $\psi(\ell^\times)$ is
related to $\log R = \ell^\times$ via Eq. (\ref{psil}))
 gave a better fit than
the derivative of the approximate expression in Eq. (\ref{log}).

For $q<4$, to leading order in $\epsilon'$, the same procedure
turns Eq. (\ref{psilq}) into
\begin{equation}
M_S \propto R^{D_S}(1-\hat B R^{-\theta})^{-c_S/a} \approx
R^{D_S}(1+f_S R^{-\theta}), \label{corr}
\end{equation}
where $D_S \approx y_S-c_S \epsilon'$ and $\theta \approx 2a
\epsilon'$. The RHS of this equation remains correct also for
higher orders in $\epsilon'$. Note that to the lowest order in
$\epsilon'$, the ratios $f_S/f_{S'}$ are universal, being equal to
$c_S/c_{S'}$. This is similar to analogous ratios for
thermodynamic properties in the usual
$\epsilon$-expansion\cite{AA}. Expanding the exact $D_S$ (Table 1)
in $\epsilon'$ yields $c_S$. Using also $a=1/\pi$ (see below)
yields our predictions for $c_S/a$ (given in Table 1), to be used
in fitting Eq. (\ref{log}). The form on the RHS of Eq.
(\ref{corr}) is already implied by den Nijs\cite{denN}, who found
that the pair correlation functions $G_H(r)$ can be expanded as a
sum over $r^{-2x_n}$, implying a leading correction exponent
$\theta=2(x_{n+1}-x_n)=4(4-g)/g$. Expanding this expression in
powers of $\epsilon'$ yields the coefficients $a=1/\pi$ and
$b=-1/2\pi$, which we use in our fits to Eq. (\ref{psil}). The
value $a=1/\pi$ also agrees with Cardy {\it et al.}\cite{cardy}.
This expression for $\theta$ also reproduces known results for
$q=2, 3$, as listed in Table 1.

\section{COULOMB GAS}
\noindent The second source of corrections involves new
contributions to the relevant pair correlation functions in the
Coulomb gas representations\cite{denN}. In some of the exact
derivations, the $q$-state Potts model renormalizes onto the
vacuum phase of the Coulomb gas, involving `particles' with
electric and magnetic `charges' $(e,m)$. At criticality, the
corresponding Coulomb gas has a basic `charge' $\phi = |2 - g/2|$
${\rm mod}~4$. Various Potts model two-point correlation functions
$G_S^P(\vec r)$ are then mapped onto Coulomb gas analogs, which
give the probability of finding two charged particles at a
distance $r$ apart. Asymptotically, these are given by
\begin{equation}
G^{CG}_{[(e_1,m_1),(e_2,m_2)]} (\vec{r}) \propto
r^{-2x^{CG}_{[(e_1,m_1),(e_2,m_2)]}},
\label{corr.eq}
\end{equation}
where
\begin{equation}
x^{CG}_{[(e_1,m_1),(e_2,m_2)]}=  - \frac{e_1 e_2}{2g} - \frac{gm_1
m_2}{2}. \label{exponent.eq}
\end{equation}
Hence one identifies $D_S=d-x^{CG}_{[(e_1,m_1),(e_2,m_2)]}$, with
$d=2$. The results in Table 1 for $S=M$ were obtained by den
Nijs\cite{denN}, who noted that the spin-spin correlation function
of the Potts model maps onto a Coulomb gas total electric charge
$Q=-2\phi$, which splits into the two charges $e_{1,2}=\pm 1-\phi$
(and $m_{1,2}=0$). Continuing along similar routes, Saleur and
Duplantier\cite{Sal87} used a mapping onto the body-centered
solid-on-solid model, requiring a vortex-antivortex pair with
$e_{1,2}=-\phi$ and $m_{1,2}=\pm 1/2$ or $\pm 1$ for the fractal
dimensions of $S=H$ or $S=SC$. The Table also contains
Duplantier's recent result\cite{Dup00} for $D_{EP}=2-x^P_{EP}$,
which has not been expressed in terms of Coulomb charges. The
results for $x^P_H$ and $x^P_{SC}$ are special cases of the
expression $x_\ell=g \ell^2/32-(4-g)^2/(2g)$, with $\ell=2$ and 4
respectively\cite{Sal87}. For percolation ($q=1$ and $g=8/3$),
this expression also yields $x^P_{EP}=x_3=2/3$ for the external
perimeter and $x^P_G=x_6=35/12$ for the gates to
fjords.\cite{Aiz99}

We now turn to corrections to the leading behavior. den
Nijs\cite{denN} derived such corrections for the order parameter
correlation function. In that case, he noted that the charge
$Q=-2\phi$ could also split into the pair $e_{1,2}=\pm 3-\phi$,
yielding a contribution to $G^P_M$ of the form $r^{-2 x^P_{M,2}}$,
with $x^P_{M,2}=x^{CG}_{[(3-\phi,0),(-3-\phi,0)]}=x^P_M+4/g$.
Since $D=d-x$ usually represents a fractal dimension, we relate
each of these correction terms to some subset of the cluster, with
dimension $D_{M,2}=2-x^P_{M,2}=D_M-4/g$. Writing $M_M$ as a sum of
powers $R^{D_i}$,\cite{partD} we have $M_M \propto
R^{D_M}(1+f'R^{-\theta'})$, with $\theta'=4/g$.

As far as we know, there has been no discussion of the analogous
corrections to the other subsets discussed here. In the spirit of
den Nijs\cite{denN}, we note that the correlation function for
both $H$ and $SC$ could also result from electrical charges
$e_{1,2}=\pm 2-\phi$, instead of $-\phi$. For both of these cases
this would give $x'=x+2/g$, hence a correction exponent
$\theta''=2/g$. At the moment, there exists no theory for
corrections to $M_{EP}$. However, in the spirit of the
renormalization group it is also reasonable to interpret $\theta'$
and $\theta''$ as the scaling exponent of some irrelevant
perturbation (yet to be identified). If that were true then we
might expect the same perturbation also to affect other
quantities, like $M_{EP}$. This conjecture is supported by the
`superuniversal' relation, $(D_H-1)(D_{EP}-1)=1/4$, found by
Duplantier\cite{Dup00}. If this relation also holds for the
effective dimensions (as happens e. g. in the
$\epsilon$-expansion\cite{AA}), then $H$ and $EP$ should have the
same correction exponents.

\section{ANALYTIC CORRECTIONS; SUMMARY}

\noindent The last source of corrections involves `analytic'
terms, coming e. g. from linear cuts with dimensions $(D_S-1)$,
\cite{partD} or from replacing $R$ by $(R+a)$, since there are
many possible candidates for the correct linear measure of the
cluster. These would imply corrections of relative size $1/R$.

Combining all of these sources, we end up with the prediction (for
$q<4$)
\begin{equation}
D_S^{\rm eff}=D_S+\sum_i f_i R^{-\theta_i},
\end{equation}
with $\theta_i= \theta,~\theta'$ (or $\theta''$) and 1. Indeed,
our numerical simulations\cite{bbm} basically confirm these
expressions.

In summary, we have presented several general expressions for the
$q$-dependent corrections to the asymptotic $R$-dependence of the
mass, hull, external perimeter and singly connected bonds. Such
corrections are crucial for fitting numerical data. It would be
nice to have a unifying theory, which would confirm these
expressions in a rigorous way. It would also be nice to obtain
similar corrections for other geometrical and physical quantities,
e. g. the number of gates to fjords\cite{Aiz99}.

\vspace{1cm}

\noindent {\bf ACKNOWLEDGMENTS}
\vspace{.3cm}

\noindent This paper is dedicated to Antonio Coniglio, on the
occasion of his 60th birthday. Coniglio contributed a lot to our
understanding of many issues discussed in this paper, both through
his publications and through very stimulating personal discussions
during the last quarter of a century. This project is part of a
larger collaboration with Benoit Mandelbrot, who initiated it and
continuously contributed via many discussions, and also with Juha
Pekka Hovi and E. Rausch. We also acknowledge support from the
German-Israeli Foundation. This work has also been supported in
part by the Academy of Finland through its Center of Excellence
program.

\newpage
\noindent {\bf REFERENCES}
\vspace{.3cm}

\end{document}